\documentstyle[graphicx,prl,aps]{revtex}
\begin{document}
\paperwidth =15cm 
\draft
\title{Comment on ``Negative heat $\cdots$'' by Moretto et al.}
\author{D.H.E. Gross}
\address{
Hahn-Meitner-Institut
Berlin, Bereich Theoretische Physik,Glienickerstr.100\\ 14109 Berlin, Germany
and 
Freie Universit{\"a}t Berlin, Fachbereich Physik; \today}
\maketitle
\begin{abstract} 

\end{abstract}
Backbending of the caloric curve $T(E)$ is experimentally found in
nuclear multifragmentation e.g.\cite{pochodzalla95,dAgostino00}.  In
their recent paper (arXiv:nucl-th/0012037) Moretto et al.  claim that
the first order transition and the backbending of the caloric curve in
small systems like fragmenting nuclei is trivial. Of course well
understood physics is always trivial, but in this case they miss some
essential facts which make phase transitions in such small systems
interesting by generic reasons.
\begin{itemize}
\item Nuclear multifragmentation is not the story of a single nuclear
  drop embedded in its vapor at some given temperature and pressure.
  In nuclear fragmentation {\em experiments} there is a constant
  energy but neither a heat bath (temperature) nor a vapor pressure. I.
  e. the realistic scenario is not as trivial as Moretto et al. claim.
\item {\em Multi}fragmentation is first of all a fast, nearly
  simultaneous but {\em statistical} breaking of a hot nucleus into
  {\em several intermediate nuclei} with several surfaces. As was
  discussed in \cite{gross161} and earlier papers, the intruder in
  $S(E)$ at the beginning of the transition is caused by the increase
  of the number $N_{fr}$ of intermediate fragments and thus of the
  {\em total} surface area including a partition entropy $S_{part}$
  connected to various partitions into several fragments with a
  fluctuating number. With no $S_{part}$ there would be no {\em
    multi}fragmentation. This aspect is lost in Moretto's scenario.
\item The Rayleigh (curvature-) pressure of a liquid drop invoked by
  Moretto has of course its microscopic origin in the surface tension
  which is a cooperation of surface energy and the {\em entropy} of
  the surface.  Without surface tension there is no Rayleigh pressure.
\item Their example of a nucleus with 20 nucleons is far away from a
  macroscopic liquid drop with a surface. There is nearly no volume
  part in such nuclei.
\item If $T$ and $P$ are fixed as one may interprete their fig.3 they
are in the \{T,P,N\}-ensemble. Then neither the overall-volume nor
the enthalpy are fixed. In the stationary approximation of the
Laplace-transform $\{H,V\}\to\{T,P\}$ (and this corresponds to
standard thermodynamics) there are two stationary points: one on the
left branch of their fig.3 and one on the right branch. The middle
downwards going branch is unstable.  I.e. the system should jump
along a horizontal line ($\sim$ Maxwell construction) from the left
branch to the right one and the backward (interesting) branch remains
unvisible.  If, however, their figure means that they fix the
enthalpy and determine the temperature implicitly by $T=(dS/dH)^{-1}$
then they are in the {\em micro-canonical} ensemble with respect $H$
which is {\em not} standard thermodynamics for a small system.
In full consistency with my general statements then the backbending
can be observed.  Thus the backbending is indeed a {\em specific feature
of microcanonicity} in sharp contrast to their concluding point 3.
\item On a much more fundamental level: Nearly all textbooks of
  statistical thermodynamics claim that phase transitions exist in the
  thermodynamic limit only.  It is an important, anything else but
  trivial, and for thermodynamics in general quite fundamental
  question whether one can define phase transitions of any order and
  (multi-)critical phenomena unambiguously in such small,
  non-extensive systems as hot nuclei. A question addressing the
  majority of systems in nature like also atomic clusters, biological
  and astrophysical systems.  This is discussed in \cite{gross174}.
\end{itemize}
The contribution of Moretto et al. is a nice exercise in macroscopic
liquid drop theory but misses the crucial features of {\em nuclear
  multi}fragmentation.  It is too trivial and moreover wrong.

\end{document}